\documentclass[12pt]{article}
\usepackage{epsfig,a4wide}

\usepackage{amsmath}
\usepackage{amssymb}
\usepackage{graphicx}
\usepackage{epsfig}
\usepackage{bm}

\newcommand{\be}{\begin{equation}}
\newcommand{\ee}{\end{equation}}
\newcommand{\ba}{\begin{eqnarray}}
\newcommand{\ea}{\end{eqnarray}}

\begin{document}

\title{Inclusive $\omega$ photoproduction from nuclei
and $\omega$ in the nuclear medium}

\author{M. Kaskulov$^1$\thanks{e-mail: kaskulov@ific.uv.es},
~E. Hernandez$^2$\thanks{e-mail: gajatee@usal.es} 
~and~ E. Oset$^1$\thanks{e-mail: oset@ific.uv.es} \\
{\small $^1$~Departamento de F\'{\i}sica Te\'orica and IFIC,
Centro Mixto Universidad de Valencia-CSIC,} \\
\small Institutos de
Investigaci\'on de Paterna, Aptd. 22085, 46071 Valencia, Spain \\
\small $^2$~Grupo de F\'{\i}sica Nuclear, Departamento de Fisica Fundamental
e IUFFyM, \\
\small Facultad de Ciencias, 
Universidad de Salamanca, \\
\small Plaza de la Merced s/n,
E-37008 Salamanca, Spain
}
%
%

\date{\today}
\maketitle

\begin{abstract}  
With the aim of extracting information on the shift of the $\omega$ mass in the
nuclear medium we analyze data obtained at ELSA from where claims for evidence
of a mass shift of the $\omega$ have been made. We develop a Monte Carlo
simulation code which takes into account the possible reactions in the
experimental set up of $(\gamma ~ A \to \pi^0 \gamma ~X)$ in the vicinity of
the $\omega$ production region with subsequent $\omega \to \pi^0 \gamma$ decay.
We compare our results with experiment for the distribution of  $\pi^0 \gamma$
invariant masses and conclude that the distribution is compatible with an
enlarged $\omega$ width of about 90 MeV at nuclear matter density and no shift
in the mass. This change in the width would be compatible with the  preliminary
results obtained from the transparency ratio in the A dependence of $\omega$
production. The discrepancy of the present conclusions with former claims of an
evidence for a shift of the $\omega$ mass stem from a different choice of
background which is discussed in the paper.  

\end{abstract}

\section{Introduction}

The interaction of mesons with nuclei has captured the attention of hadron
community and much work has been done on the topic~\cite{Post:2003hu}.
In particular the behavior of vector mesons
in nuclei has been thoroughly studied,
stimulated by the ansatz of a universal scaling of the vector meson masses in
nuclei suggested in \cite{scaling} and the study of QCD sum rules  in
nuclei \cite{hatsuda}. Although most of the efforts have been devoted to the
change of the $\rho$ properties in the medium, the properties of the 
$\omega$ meson have also received much attention and many theoretical efforts
have been devoted to obtain the changes of the mass and width  in the medium 
\cite{jean,klingl1,saito,tsushima,friman,klingl2,post,saito2,lykasov,sibirtsev,dutt,lutz,zschocke,dutt2,muhlich,mosel,muhlich2,steinmueller}. 
 The values obtained for the selfenergy of 
the $\omega$ in nuclei split nearly equally into attraction and repulsion 
and range from an attraction of the order of 100-200 MeV 
\cite{tsushima,klingl2} to no changes in the mass \cite{mosel} to a net
repulsion of the order of 50 MeV \cite{lutz}.

 While most of the experimental work is conducted in heavy ion reactions, 
 it has been argued in \cite{moseltalk} that reactions
involving the interaction of elementary particles with nuclei can be equally
good to show medium effects of particles, with the advantage of being easier to
analyze. In this sense, a variety of experiments have been done with $pA$
collisions in nuclei at KEK \cite{ozawa,tabaru,sakuma} and photonuclear collisions at
Jefferson lab \cite{weygand} by looking at dilepton spectra. 

   A different approach has been followed by the CBELSA/ TAPS collaboration by
   looking at the $\gamma \pi ^0$ coming from the $\omega$ decay. 
In this line a
    recent work \cite{trnka}
claims evidence for a decrease of the $\omega$ mass in the medium of the
order of 100 MeV from the study of the modification of the mass
spectra in $\omega$ photoproduction.

In the present work we perform a reanalysis of the data of \cite{trnka}. 
We develop a Monte Carlo
simulation code which takes into account the possible reactions in the
experimental set up of $(\gamma ~ A \to \pi^0 \gamma ~X)$ in the vicinity of
the $\omega$ production region with subsequent $\omega \to \pi^0 \gamma$ decay.
Especial emphasis is done in the final state interaction of the particles
produced in order to properly reconstruct the invariant mass of the   $\pi^0
\gamma$ subsystem.  We first look at the A dependence of the $\omega$
production (transparency ratio) from where, in base to preliminary data, we
induce an approximate  width of the $\omega$ in the nuclear medium. Next we
compare our results with experiment for the distribution of  $\pi^0
\gamma$ invariant masses and conclude that the distribution is compatible with
an enlarged $\omega$ width of about 90 MeV at nuclear matter density and no
shift in the mass. The discrepancy of the present conclusions with former
claims of an evidence for a shift of the $\omega$ mass stem from a different
choice of background. We show that the former claims were based on an 
assumption of background which eliminates strength at high $\omega$ masses,
while the choice of background scaled from the elementary reaction
automatically leads to strength at high $\omega$ masses, as well as in lower
masses, which can be explained by means of an increased $\omega$ width, also
compatible with the preliminary results obtained from the transparency ratio.

The paper proceeds as follows: In Section 2 we outline the results of the
microscopic calculations which serves here as input for our
computer simulation. In Sections 3 and 4 
we provide the details of the Monte Carlo simulation method.
The results are presented in Sections 5,6 and 7. The conclusions are given
in Section 8.


\section{Preliminaries}

The inclusive reactions like the one studied here, 
$A(\gamma,\omega\to\pi^0\gamma) X$, 
require minimum knowledge of nuclear structure,
since one is not looking at any particular final
nuclear state but integrating over all of them. 
Assuming further, that the elementary production amplitude on a single 
nucleon has been fixed, in a first step,
the nuclear cross section can be obtained by summing  over the occupied 
nuclear states in the Fermi sea and by
introducing the standard nuclear 
effects like the Fermi motion of the initial
nucleons and the Pauli blocking (for the $\gamma N \to \omega N$)
of the outgoing ones. 

In the present work we go beyond this scheme and 
consider also the final state interaction (FSI) 
of the $\omega$-mesons and its decay
products 
in finite nuclei. 
The method used here will 
combine a phenomenological 
calculation of the intrinsic probabilities for different nuclear 
reaction, like the quasielastic and absorption channels, as a function of
the nuclear matter density, 
followed by a computer Monte Carlo (MC) simulation procedure in order to 
trace the fate of the $\omega$-mesons and its 
decay products in the nuclear medium.
Since our calculations represent complete event simulations
it will be possible to account for the actual experimental 
acceptance effects. 
In the following we shall carry out
the computer MC simulation taking into account the  
geometrical and kinematical acceptance conditions of the 
 Crystal Barrel/TAPS 
experiment at ELSA. 

We consider the  photonuclear reaction 
$A(\gamma,\omega\to \pi^0\gamma)X$ in two
steps - 
production of the $\omega$-mesons and 
propagation of the final states.
In the laboratory, where 
the nucleus with the mass number $A$ is at rest, 
the nuclear total  cross section  of the inclusive reaction $A(\gamma,\omega)X$
including the effects of Fermi motion and Pauli blocking is given by
\begin{eqnarray}
\label{TCSnuclear}
&& \sigma_{\gamma A\to \omega X} = \nonumber \\ 
&& \int_{A}   
\frac{d^3 \vec{r}}{2\pi^2} \int \frac{d^3\, \vec{p}_N}{(2\pi)^3} 
\int d\widetilde{m}_{\omega}^2 \, 
\frac{S_{\omega}({m}_{\omega},\widetilde{m}_{\omega},\rho_A(r)) M_N^2}
{ 
|\, \vec{p}_{\gamma} + \vec{p}_{N}|
[(p_{\gamma}+p_N)^2-M_N^2]}
 \nonumber \\
&& \times  \int 
dE_{\omega} 
\int_{0}^{2\pi} d\varphi_{\omega} 
\Big(\overline{\sum_{\mbox{\tiny in }}} \sum_{\mbox{\tiny out }} 
T^{\dagger} T \Big)_{\gamma N \to \omega N} \, \Theta(1-B^2) \nonumber \\
&& \times \, \Theta(k_F(r)-|\vec{p}_N|) \,   
\Theta(E_{\gamma} + E_N - E_{\omega} -E_F(r)) 
\nonumber \\
&& \times \,
\Omega_{\omega}(\vec{r},\rho(r),\vec{p}_{\omega},\widetilde{m}_{\omega}) 
\end{eqnarray}
Here 
$\Theta$ denotes the step function and
$B$ stands for the cosinus of the angle between 
$\vec{p}_{\gamma}+\vec{p}_{N}$ 
and $\vec{p}_{\omega}$ 
\begin{eqnarray}
B &\equiv& \cos\vartheta_{\omega} = 
 \frac{1}{2\, |\vec{p}_{\omega}||
\vec{p}_{\gamma}+\vec{p}_{N}
|}  \\
&& \times \left[ 
(\vec{p}_{\gamma}+\vec{p}_{N})^2 +\vec{p}_{\omega}^{\,2}+ M^2_{N} 
- \left( E_{\gamma} + E_N - E_{\omega}\right)^2 \right] \nonumber
\end{eqnarray}
with obvious notations for the momenta and
energies of the particles $E_{\gamma}= |\vec{p}_{\gamma}|$,
$E_{\omega}=\sqrt{\vec{\, p}_{\omega}^2+\widetilde{m}_{\omega}^2}$
and $E_{N}=\sqrt{\vec{\, p}_{N}^2+M_{N}^2}$. 
Also in Eq.~(\ref{TCSnuclear}) $E_F(r) = \sqrt{ k_F^2(r)+M_N^2}$ 
is the local Fermi energy and
the Fermi momentum $k_F$ is related
to the local
density $\rho_A(r)$ of the nucleus by
\begin{equation}
\rho_A(r) = 4 \int \frac{d^3\vec{p}_N}{(2\pi)^3} \Theta(k_{F}(r)-|\vec{p}_N|)
= \frac{2\, k_F^3(r)}{3\pi^2}.
\end{equation}

\noindent 
The photoproduction amplitude entering Eq.~(\ref{TCSnuclear}) is that from
the elementary reaction $N(\gamma,\omega)N$  properly summed
and averaged over the final and initial polarizations, respectively. 
It is given by 
\begin{equation}
\label{ElemAmplitude2}
\overline{\sum_{\mbox{\tiny in }}} 
\sum_{\mbox{\tiny out }} 
T^{\dagger} T 
= \frac{16\pi s E_{\gamma}^{*2}}{M^2_N} 
\frac{d\sigma_{\gamma N\to \omega N}}{dt}
\end{equation}
where $d\sigma_{\gamma N\to \omega N}/dt$ 
is an invariant differential cross section with 
$s=(p_{\gamma}+{p}_N)^{\,2}$, 
$E_{\gamma}^* = \frac{s-M_N^2}{2\sqrt{s}}$ and 
$t=(p_{\omega}^{\,lab}-p_{\gamma})^2$ where $p_{\omega}^{\,lab}$ is the
 four momentum of the $\omega$ in the laboratory frame.
In Eq.~(\ref{TCSnuclear}) the  momentum 
$\vec{p}_{\omega}=(|\vec{p}_{\omega}|,\theta_{\omega},\phi_{\omega})$ 
of the $\omega$ is 
defined with respect to 
$\vec{p}_{\gamma}+\vec{p}_N = 
(|\vec{p}_{\gamma}+\vec{p}_N|,\theta',\phi')$.
Making use of ordinary rotation
matrices $\mathcal{R}_{\varphi'}$ and
$\mathcal{R}_{\theta'}$ one can transform 
$\vec{p}_{\omega}$ 
to the laboratory system where it takes the form
\begin{equation}
\vec{p}_{\omega}^{\,lab} = \mathcal{R}_{\varphi'} \otimes
\mathcal{R}_{\theta'}  \otimes \vec{p}_{\omega}
\end{equation}
with
\begin{equation}
\mathcal{R}_{\phi'} = 
\left(
\begin{array}{ccc}
\cos \varphi' &-\sin \varphi'&0 \\
\sin \varphi' & \cos \varphi'&0 \\
0             &0             &1
\end{array}
\right), ~
\mathcal{R}_{\theta'} = 
\left(
\begin{array}{ccc}
\cos \vartheta' &0             &\sin \vartheta' \\
0               &1             &0 \\
-\sin \vartheta'&0             &\cos \vartheta'
\end{array}
\right).
\end{equation}

\begin{figure}[t]
\begin{center}
\includegraphics[clip=true,width=0.7\columnwidth,angle=0.]
{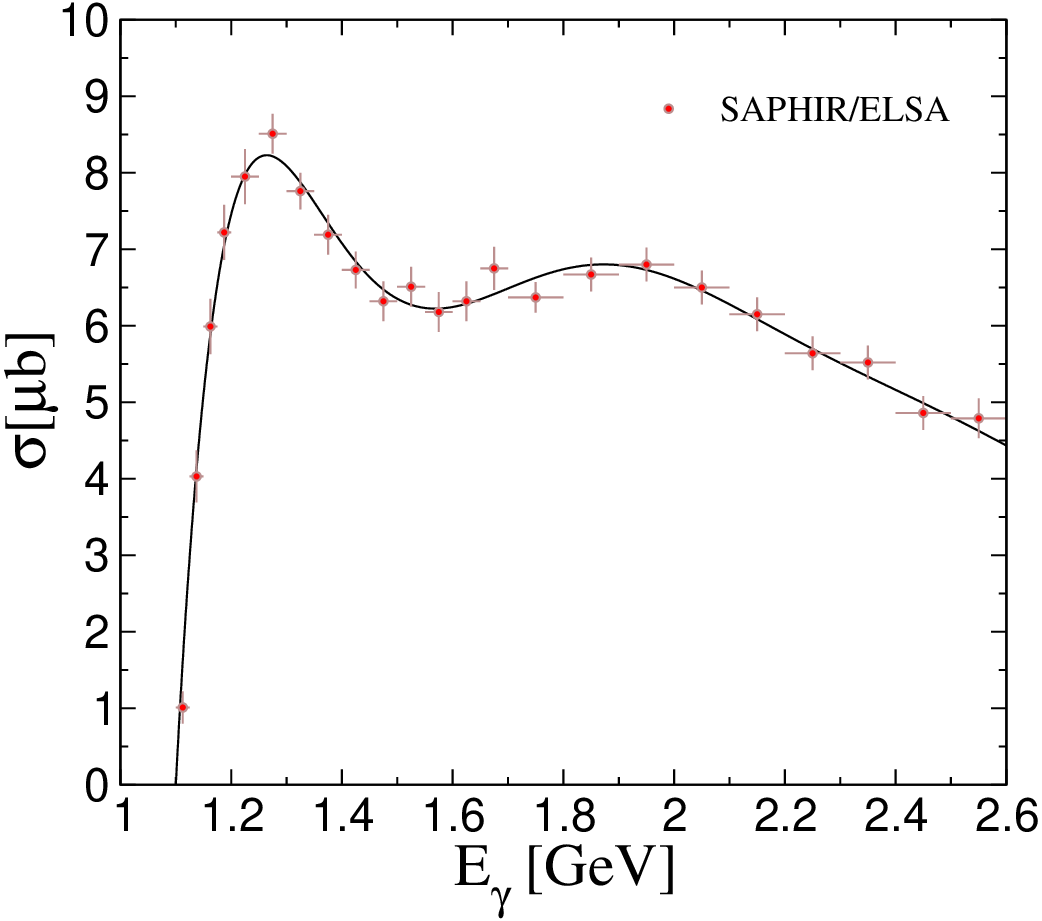}
\caption{\label{Figure1} \footnotesize
Total cross section of the reaction $p(\gamma,\omega)p$
   as a function of the photon energy $E_{\gamma}$. 
The experimental data
are from Ref.~\cite{Barth}.  The solid curve 
is the fit to data.}
\end{center}
\end{figure}

In Ref.~\cite{Barth} 
$d\sigma_{\gamma p \to \omega p}/dt$  of the reaction 
$(\gamma,\omega)$ on hydrogen target
followed by the $\omega \to \pi\pi\pi$ decay has been measured
for incident photon energies from the reaction
threshold $E_{\gamma}^{th} = m_{\omega}+m_{\omega}^2/2M_p \simeq 1.1$~GeV 
up to 2.6~GeV. In the present work the data of Ref.~\cite{Barth} 
are conveniently parameterized
and properly implemented in the MC code. The fit to  
the total cross section
is presented in Fig.~\ref{Figure1} where
the experimental data from Ref.~\cite{Barth} are also shown.
In the following, 
the cross section on the neutron will be taken to be the same as on a proton.

The $\omega$-mesons are produced according to their
spectral function $S_{\omega}$ at a local density $\rho(r)$
\begin{eqnarray}
\label{SF}
S_{\omega}({m}_{\omega},\widetilde{m}_{\omega},\rho) = 
\hspace{5cm}\nonumber \\
- \frac{1}{\pi} 
\frac{\mbox{Im}\Pi_{\omega}(\rho)}
{\Big(\widetilde{m}_{\omega}^2-{m}_{\omega}^2
-\mbox{Re}\Pi_{\omega}(\rho)\Big)^2 + 
\Big(\mbox{Im}\Pi_{\omega}(\rho)\Big)^2},
\end{eqnarray}
where $\Pi_{\omega}$ is the in-medium selfenergy of the 
$\omega$ with nominal mass $m_{\omega}=782$~MeV. The width of the 
$\omega$ in the nuclear medium is 
related to the selfenergy by 
$\Gamma_{\omega}(\rho,\widetilde{m}_{\omega}) 
= - \mbox{Im}\Pi_{\omega}(\rho,\widetilde{m}_{\omega})/E_{\omega}$.
It includes
the free width $\Gamma_{free} = 8.49$~MeV and an in-medium 
part $\Gamma_{coll}(\rho)$ which accounts for the 
collisional broadening of the $\omega$ due to the quasielastic and
absorption channels to be specified below. In Eq.~(\ref{SF})
$\mbox{Re}\Pi_{\omega}=2 E_{\omega} \mbox{Re}V_{opt}(\rho)$, where
$V_{opt}(\rho)$ is the $\omega$ nucleus optical potential accounts 
for a possible
shift of the $\omega$ mass in the medium and we shall make some considerations
about it latter on.

In Eq.~(\ref{TCSnuclear})  
the distortion factor $\Omega_{\omega}$ 
describes the propagation and decay of the
$\omega$ mesons inside and outside the nucleus 
as well as the FSI of its decay products - in our case $\pi^0 \gamma $ - 
in the nuclear medium. 
It is the subject of the present
MC simulation.

In the following we shall also consider the situation when 
the energy of the incident photon beam  is not fixed but 
constrained in some energy interval
$E_{\gamma}^{\min} < E_{\gamma} < E_{\gamma}^{\max}$.
This condition is implemented by folding the
cross section of Eq.~(\ref{TCSnuclear})
with the photon
flux profile 
\begin{equation}
\label{TCSnuclearBr}
\int_{E^{\min}_{\gamma}}^{E^{\max}_{\gamma}} 
dE_{\gamma} \, W_{\gamma}(E_{\gamma}) \, \sigma_{\gamma A \to \omega X}
\end{equation}
At ELSA facility, the photon beam is produced via the electron bremsstrahlung
and therefore one creates the photon flux according to the unnormalized
spectra
\begin{equation}
W_{\gamma}(E_{\gamma}) = \frac{1}{N_{\gamma}} \frac{dN_{\gamma}}{dE_{\gamma}}
\propto \frac{1}{E_{\gamma}}.
\end{equation}

\section{The Monte Carlo simulation procedure}

The computer MC simulation proceeds in close analogy to the actual
experiment.
At first,
the multiple 
integral in Eq.~(\ref{TCSnuclear}) is carried out using 
the MC integration 
method. This procedure provides
 a random 
point $\vec{r}$ inside the nucleus where the photon collides 
with the nucleon, 
also randomly generated
from the Fermi sea with $|\vec{p}_N| \le k_F(\vec{r})$, 
see the factor $\Theta(k_F(\vec{r})-|\vec{p}_N|)$ 
in Eq.~~(\ref{TCSnuclear}).  
For the sample event in the MC integral
the mass $\widetilde{m}_{\omega}$ of the $\omega$
respects 
the spectral function $S_{\omega}$ at local density $\rho(r)$, 
see Eq.~(\ref{SF}).
Inside the nucleus the $\omega$-mesons moving with the three
momentum $\vec{p}_{\omega}^{\,lab}$ necessarily interact with
the nucleons in their way out of the nucleus. In the MC simulation
the $\omega$-mesons are allowed to propagate  
a distance 
$\delta \vec{L} = \frac{\vec{p}_{\omega}^{\,lab}}{|\vec{p}_{\omega}^{\,lab}|}
\delta L$ and at each step, $\delta L \simeq 0.1$~fm, the 
reaction probabilities for different 
channels like
the decay of the $\omega$
into $\pi^0 \gamma$ and $\pi\pi\pi$ final states, 
quasielastic scattering and in-medium absorption
are properly calculated. 

The conventional decay channels are  
$\omega \to \pi\pi\pi$ and $\omega \to \pi^0\gamma$ 
and the corresponding reaction probabilities per unit length are given by 
\begin{equation}
\frac{\delta P_{\omega\to \pi\pi\pi}}{\delta L} 
= \frac{1}{\gamma v} \, \Gamma_{\omega\to \pi\pi\pi} = 
\frac{m_{\omega}}{|\vec{p}_{\omega}|}  \, \Gamma_{\omega\to \pi\pi\pi} 
\end{equation}

\begin{equation}
\frac{\delta P_{\omega\to \pi\gamma}}{\delta L} = 
\frac{1}{\gamma v} \, \Gamma_{\omega\to \pi\gamma} 
= \frac{m_{\omega}}{|\vec{p}_{\omega}|} 
\, \Gamma_{\omega\to \pi\gamma} 
\end{equation}
where $v = |\vec{p}_{\omega}|/E_{\omega}$ is the $\omega$ velocity  and
the Lorentz contraction factor $\gamma = E_{\omega}/m_{\omega}$ 
relates the width
of the $\omega$ in the rest frame, $\Gamma_{\omega}$, to that in the moving 
frame $\Gamma^*_{\omega}$. The partial decay width 
of the $\omega$ into $\pi\pi\pi$ and $\pi^0\gamma$ decay channels are 
$\Gamma_{\omega\to \pi\pi\pi} \simeq 7.56$~MeV and
$\Gamma_{\omega\to \pi\gamma}  \simeq 0.76$~MeV, respectively~\cite{PDG}.

At given local density $\rho(r)$, 
the probability per unit length for the quasielastic collision 
of the $\omega$ is given by
\begin{equation}
\frac{\delta P_{CB}}{\delta L} 
= \sigma_{\omega N \to \omega N} \, \rho(r) 
\end{equation}
where $\sigma_{\omega N \to \omega N}$ stands for the elastic
$\omega N \to \omega N$ cross section. In the present work we employ
the parameterization of the $\sigma_{\omega N \to \omega N}$
used in Refs.~\cite{lykasov,muhlich2}.
For the sample quasielastic event the
 angular distributions (we assume $s$-wave) are generated in the c.m. frame 
of the $\omega$ and a random nucleon in the Fermi sea. 
Then a Lorentz boost
back to the laboratory is done in order to obtain the energy and 
momentum of the scattered
$\omega$ and outgoing nucleon after the quasielastic step. 
Since the outgoing
 nucleon moving with the three momentum $|\vec{p}_N'|$ 
is subject to Pauli blocking, we require that
the quasielastic scattering fulfills the condition
$|\vec{p}_N'| > k_F(r)$.

The in-medium $\omega N \to \omega N$ elastic scattering
does not lead to a loss of flux and does not change the total nuclear cross
section.  It affects the $\omega$ energy and momentum distributions only
increasing the
energy loss of the $\omega$ in MC steps. 
The loss of $\omega$ flux is related to the absorptive
part of the 
$\omega$-nucleus optical
potential. 
In nuclear matter the $\omega$ wave propagating through 
will acquire the phase 
$\sim \exp (-i V_{opt}^{abs}t)$ which enforce the $\omega$ to be 
removed from the
elastic flux at the rate
\begin{equation}
\frac{1}{N_\omega} \frac{dN_{\omega}}{dt} 
\equiv \frac{\delta P_{abs}}{dt} =
\Gamma_{abs} = -2 \, \mbox{Im}\, V_{opt}^{abs}
\end{equation}
where $V_{opt}^{abs}$ is the part of the $\mbox{Im}\, V_{opt}$
related to the the $\omega N$ inelastic cross section $\sigma_{in}$
and other many-body absorption mechanisms.

The reaction probability for the $\omega$ meson to be absorbed 
inside the nucleus after the propagation
of the length interval $\delta L$ or undergo inelastic collision is given by
\begin{equation}
\frac{\delta P_{abs}}{\delta L} =  \frac{1}{v} \Gamma_{abs}
\end{equation}
 As a first estimate we use the following parameterization for the width
\begin{equation}
\label{Gabs}
\Gamma_{abs} = \Gamma_0 \frac{\rho(r)}{\rho_0}
\end{equation}
where $\rho_0=0.16$~fm$^{-3}$ is the normal nuclear matter density.
The parameterization is adequate for the inelastic processes and only
approximate for the absorption processes which would be better represented by
a $\rho^2$ functional. It is thus implicit that the $\rho$ functional for the
absorption provides the correct absorption width for the average density felt
by the process.

Since we are interested in $\pi^0\gamma$ events,
the absorption channels and decay $\omega \to \pi\pi\pi$ 
remove the $\omega$-mesons from initial flux.
The $\pi^0 \gamma$ events may come from both the $\omega$
decaying inside and outside the nucleus. Only $\pi^0\gamma$ events
from $\omega\to \pi^0 \gamma$ 
decays inside the nucleus carry information on the
$\omega$ in-medium properties.
If the resonance leaves the nucleus, its spectral function must coincide with
the free distribution,  
$\mbox{Im}\Pi_{\omega} = - \widetilde{m}_{\omega} \Gamma_{\omega}^{free}$, 
because the collisional part of the width is zero in
this case.
When the $\omega$ decays into $\pi^0 \gamma$ pair 
at point $\vec{r}'$ inside the nucleus 
its mass distribution is generated  according to the 
in-medium spectral function at the local density $\rho(r')$. 
For a given mass $\widetilde{m}_{\omega}$ 
the $\omega$-mesons are allowed to decay 
isotropically in the c.m. system 
into the $\pi^0 \gamma$ channel. 
 The direction of the $\pi^0$
(therefore $\gamma$)
is then chosen randomly  and an appropriate Lorenz transformation is done
in order to generate the corresponding $\pi^0 \gamma$ distributions in the
laboratory frame. 
 The
$\omega$-mesons are reconstructed using the energy and 
momentum of the $\pi^0 \gamma$
pair in the laboratory.


\section{Propagation of pions in nuclei}
The reconstruction of the genuine
$\omega{\to}\pi^0\gamma$ mode is affected by the FSI of the $\pi^0$  
in the nucleus which distorts the
$\pi^0 \gamma$ spectra. 
In this case, if the $\pi^0$ events come from the interior of the nucleus 
we trace the fate of the neutral pions 
starting  
from the decay point of the $\omega$-meson.
 In their way out
of the nucleus pions can experience the quasielastic scattering or 
can be absorbed. 
The intrinsic probabilities for these reactions
as a function of the nuclear matter density
are calculated using the phenomenological models of 
Refs~\cite{Salcedo:1987md,Oset:1986sy,Oset:1990zj},
which also include 
higher order quasielastic cuts and 
the two-body and three-body
absorption mechanisms.
Since the FSI of the $\gamma$ quanta 
are rather weak they  are allowed to escape the nucleus without distortion.
In this section we briefly summarize the approach used
for the description of the
pion propagation in nuclei. 

We consider different energy regions where pions
fall according to their kinetic energy $T_{\pi}$. 
In the $\Delta(1232)$ 
region 
we use the microscopic $\Delta$h model for the pion nuclear 
interaction~\cite{Salcedo:1987md,Oset:1986sy}.
The formulae for the  probabilities per unit length that a pion
undergoes quasielastic scattering $P_{QE}$ 
(for $T_{\pi} \le 390~\mbox{MeV}$)
and that the pion is absorbed $P_A$ 
(for $T_{\pi} \le 315~\mbox{MeV}$) 
are adopted from Ref.~\cite{Hernandez:1986ng}.
The later includes higher order quasielastic cuts and 
also the parts corresponding
to the two-body and three-body
absorption cuts.


Beyond the $\Delta(1232)$-isobar region, we rely upon elementary 
$\pi$-nucleon cross sections, which provide the probability per unit length
for a certain reaction to happen.
For instance, in that region 
the probability that a pion 
undergoes a quasi-elastic scattering $P_Q$ is given by 
\begin{equation}
\label{Pr}
\frac{\delta P_{QE} 
}{\delta L}  =  
\bar\sigma_{QE} 
\rho(r)
\end{equation}
where $\bar \sigma_{QE}$ is an average of the $\pi^0p \to \pi^0p$ 
and $\pi^0n \to \pi^0n$ cross
sections. It can be determined using the isospin formalism in terms of
the experimentally accessible cross sections taken from Ref.~\cite{SAID}
\begin{eqnarray}
\bar \sigma_{QE} &\equiv& \frac{1}{2}
(\sigma_{\pi^0 p \to \pi^0 p} + \sigma_{\pi^0 n \to \pi^0 n}) 
\\ 
&& = \frac{1}{2}
(\sigma_{\pi^+ p \to \pi^+ p} + \sigma_{\pi^- p \to \pi^- p}
-\sigma_{\pi^- p \to \pi^0 n}). \nonumber
\end{eqnarray}
For the given quasielastic event the
 angular distributions are generated in the c.m. frame 
of the pion and a random nucleon in the Fermi sea. Then a Lorentz boost
back to the laboratory is done in order to obtain the energy and 
momentum of the
pion and outgoing nucleon after the quasielastic step. 
Since the outgoing
 nucleon moving with the three momentum $|\vec{p}_N'|$ 
is subject to Pauli blocking, we require that
the quasielastic scattering fulfills the condition
$|\vec{p}_N'| > k_F(r)$.

Making use of the isospin amplitudes we get for 
the charge exchange reaction the following cross section
\begin{equation}
\label{CX}
\bar{\sigma}_{CX} \equiv \frac{1}{2}
(\sigma_{\pi^0 n \to \pi^- p} + \sigma_{\pi^0 p \to \pi^+ p}) =
\sigma_{\pi^- p \to \pi^0 n}.
\end{equation}
And using similar isospin arguments the total reaction
cross section may be written in the form
\begin{eqnarray}
\label{TR}
\bar{\sigma}_R &\equiv& 
\frac{1}{2}(\sigma_{\pi^0 p\to X}+\sigma_{\pi^0 n\to X}) \nonumber \\
&&= \frac{1}{2}(\sigma_{\pi^- p\to X} +\sigma_{\pi^+ p\to X}). 
\end{eqnarray}
The corresponding probabilities are calculated using the expression similar
to Eq.~(\ref{Pr}). 
The reaction mechanisms corresponding to Eqs.~(\ref{CX}) and~(\ref{TR})
remove the pions from initial flux.
Finally, following the steps and parameterizations 
of Ref.~\cite{Oset:1990zj} we get
the extrapolation of the two- and three-body absorption mechanisms 
to higher kinetic energies of the pions.


\begin{figure*}[t]
\begin{center}
\includegraphics[clip=true,width=0.49\columnwidth,angle=0.]
{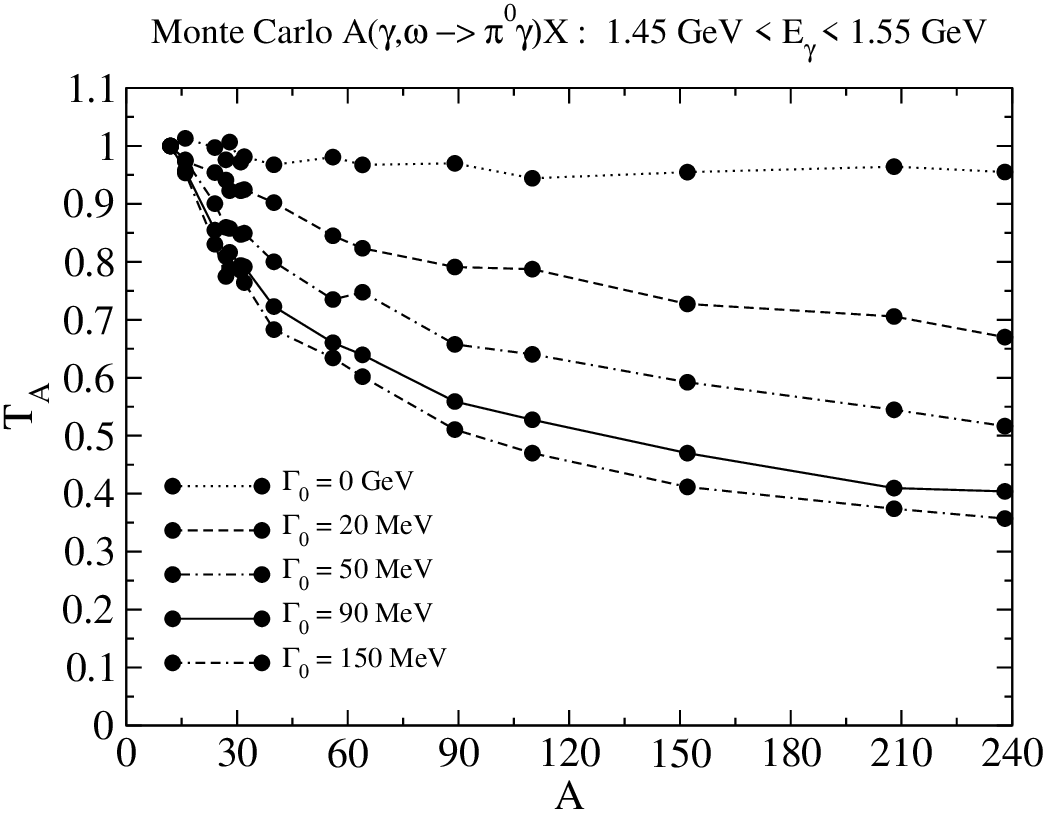}
\includegraphics[clip=true,width=0.49\columnwidth,angle=0.]
{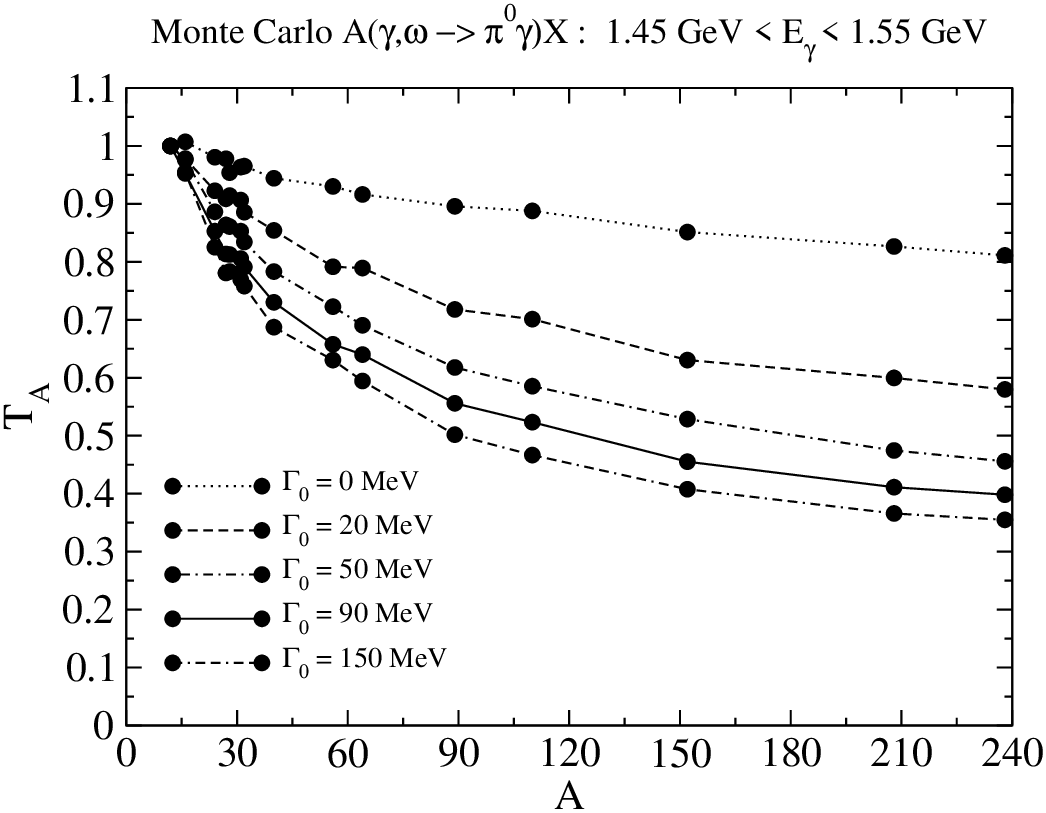}
\caption{\label{FigTrMC}  \footnotesize
The result of the Monte Carlo method for the $A$-dependence
of the nuclear transparency ratio $T_A$ without (left panel) and with
(right panel) FSI of outgoing pions.
The incident photon beam was constrained in the range
$1.45~\mbox{GeV}< E_{\gamma} < 1.55~\mbox{GeV}$.
 The carbon $^{12}$C
was used as the reference target in the ratio of the nuclear cross sections.
With
$\Gamma_{abs} = \Gamma_0 \frac{\rho(r)}{\rho_0}$ where $\rho_0$ is 
the normal nuclear matter density the dotted, dashed, dash-dotted,
solid and dash-dash-dotted  
curves 
corresponds to $\Gamma_0 = 0$~MeV, $\Gamma_0=20$~MeV, $\Gamma_0=50$~MeV, 
$\Gamma_0=90$~MeV and $\Gamma_0=150$~MeV, respectively. 
}
\end{center}
\end{figure*}

\section{In-medium $\omega$-meson width and nuclear transparency}

In this section we discuss an extraction of the 
in-medium inelastic width of the
$\omega$ in the photonuclear experiments.
As a measure for the $\omega$-meson width in nuclei we employ the so-called
nuclear transparency ratio
\begin{equation}
\tilde{T}_{A} = \frac{\sigma_{\gamma A \to \omega X}}{A \sigma_{\gamma N \to \omega X}}
\end{equation}
i.e. the ratio of the nuclear $\omega$-photoproduction cross section
divided by $A$ times the same quantity on a free nucleon. 
$\tilde{T}_A$ describes
the loss of flux of $\omega$-mesons in the nuclei and is related to the
absorptive part of the $\omega$-nucleus optical potential and thus to the
$\omega$ width in the nuclear medium. Furthermore,
the $A$ dependence of the nuclear transparency ratio
should reflect the modification of the $\omega$ meson width with increasing 
nuclear matter density. This method 
has been already proven to be very efficient in the study of the in-medium
properties of the vector mesons~\cite{magas,Muhlich:2005kf} and 
hyperons~\cite{Kaskulov:2006nm}. In Ref.~\cite{muhlich2} 
the transparency ratio has been
already used to determine the width of the $\omega$-meson in finite nuclei
using BUU transport approach.

We have done the MC calculations for the
sample nuclear targets:   ${}^{12}_6$C, ${}^{16}_{8}$O,
${}^{24}_{12}$Mg,  ${}^{27}_{13}$Al, ${}^{28}_{14}$Si,
${}^{31}_{15}$P,   ${}^{32}_{16}$S,  ${}^{40}_{20}$Ca,
${}^{56}_{26}$Fe, ${}^{64}_{29}$Cu,  ${}^{89}_{39}$Y, 
${}^{110}_{48}$Cd,  ${}^{152}_{62}$Sm,  ${}^{208}_{82}$Pb,
${}^{238}_{92}$U. 
 In the following we evaluate the ratio between the nuclear 
cross sections in
heavy nuclei and a light one, for instance  $^{12}$C, since in
this way, many other nuclear effects not related to the
absorption of the $\omega$ cancel in the
ratio~\cite{magas}. We call this ratio $T_A$.

The results of the MC calculation
for the $A$-dependence
of the nuclear transparency ratio $T_A$ are presented in Fig.~\ref{FigTrMC}.
The incident photon beam was constrained in the range
$1.45~\mbox{GeV}< E_{\gamma} < 1.55~\mbox{GeV}$ - a region which is considered
in the analysis of the CBELSA/TAPS
experiment~\cite{trnkathesis,Kotulla:2006wz}. 
The carbon $^{12}$C
was used as the reference target in the ratio of the nuclear cross sections.
In Fig.~\ref{FigTrMC} (left panel) we show the results for the
transparency ratio when the collisional broadening and FSI of the $\omega$
are taken into account but without FSI
of the pions from $\omega \to \pi^0 \gamma$ decays inside the nucleus.
With
$\Gamma_{abs} = \Gamma_0 \frac{\rho(r)}{\rho_0}$  the dotted, dash-dotted,
solid and dash-dash-dotted  
curves in Fig.~\ref{FigTrMC} (left panel)
correspond to $\Gamma_0 = 0$~MeV (no $\omega$ absorption), 
$\Gamma_0=20$~MeV, $\Gamma_0=50$~MeV, 
$\Gamma_0=90$~MeV and $\Gamma_0=150$~MeV, respectively. 
In a case without $\omega$-absorption 
we do not observe any significant decrease of the $T_A$.
Furthermore, assuming that the distortion factor in Eq.(\ref{TCSnuclear}) is 
$\Omega_{\omega}=1$ (no FSI at all)
leads essentially to the same result. For other values of the absorption
parameter $\Gamma_0$
a very strong attenuation of the
$\omega\to \pi^0 \gamma$ signal with increasing  nuclear mass number $A$
is noted.
 This is
primary due to the stronger absorption of the $\omega$-mesons 
with increasing nuclear matter density, see Eq.~(\ref{Gabs}).
Also the contribution of 
the $\omega$-mesons decaying inside the nucleus 
is increasing as a function of mass number 
$A$ merely due to an increase in the effective radius of the nucleus.

We have already noted that the FSI of the pions ($\pi$-FSI) distorts the
$\pi^0 \gamma$ spectra and the reconstructed $\pi^0\gamma$ pairs
contain events from the quasielastic steps 
which basically lose all information 
about their source. These $\pi^0\gamma$ pairs do not go into 
the final detection 
channel since they appear at much smaller 
invariant masses. 
It was already demonstrated in 
Refs.~\cite{Messchendorp:2001pa,muhlich} that the contributions
of the distorted events due to the FSI of the pions
can be largely suppressed by using a lower cut $T_{\pi} > 150$~MeV 
on the kinetic energy of 
the outgoing pions. A typical $\pi^0$ kinetic energy in the process is
$T_{\pi} \simeq 380$~MeV, hence removing pions with $T_{\pi}<150$~MeV
does indeed eliminate the pions which certainly underwent some quasielastic 
collisions.

We therefore use this cut in the full MC simulation
and, next, we consider the situation when 
both $\omega$ and $\pi^0$ are subject to FSI 
in their way out of the nucleus. 
The results of the MC simulation with the $\omega/\pi$-FSI and 
a cut $T_{\pi}>150$~MeV 
are shown in 
Fig.~\ref{FigTrMC} (right panel). As one can, see the effect of 
the $\pi$-FSI on $T_A$
is sizable at small values of the absorption parameter $\Gamma_0$. 
Note that a decrease of the ratio $T_{A}$ at $\Gamma_0=0$~MeV
is caused both, by the stronger $\pi$-absorption at higher nuclear 
matter densities and because of the cut we have imposed to remove 
the pions interacting via quasielastic scattering from the total flux.
But since the dependence of the transparency ratio on $\omega$ width
is non-linear,  the impact of the $\pi$-FSI
is already very small at $\Gamma_0 \simeq 90$~MeV. At this value of the
$\omega$-width the two curves with (right panel)
and without (left panel) $\pi$-FSI are very close to each other.


\section{The eikonal (Glauber) approximation}
In the following we  calculate the nuclear transparency ratio 
and the distortion factor 
due to the $\omega$ absorption
using the eikonal (or Glauber) approximation. 
In this framework 
the propagation of the $\omega$ in its way out of the nucleus can
be accounted for by means of 
the exponential factor
describing
the probability of loss of flux per unit
length. This simple but rather reliable method will allow us to get 
an accurate result for the integrated cross sections 
without performing an elaborate MC simulation.


We proceed as follows: let $\Pi_{\omega}$ be
the $\omega$ selfenergy in the nuclear medium
 as a function 
of the nuclear density, $\rho(r)$. We have
for the collisional width
\begin{equation}
\frac{\Gamma_{\omega}}{2} = - \frac{\mbox{Im}\Pi_{\omega}}{2 E_{\omega}}; 
\qquad
\Gamma_{\omega} \equiv\frac{d{P}}{dt} \ ,
\end{equation}

\noindent
where  $P$ is the probability of $\omega$ interaction in the
nucleus, including $\omega$ quasielastic collisions and 
absorption channels. We shall not consider the part of the 
$\mbox{Im} \Pi_{\omega}$ 
due to
the quasielastic collisions since, even if the nucleus gets excited, the
$\omega$ will still be there to be observed. Thus, mainly the absorption
of the $\omega$ is reflected in the loss of $\omega$ events
in the nuclear production as we have already demonstrated
using the MC method. 
This part of  the $\omega$ selfenergy is the one discussed before,
see Eq.~(\ref{Gabs}). Hence, we have for the probability of loss of
flux per unit length
\begin{equation}
\frac{dP}{dl}=\frac{dP}{v\,dt}
=\frac{dP}{\displaystyle \frac{|\vec{p}_\omega|}{E_\omega} dt}
=\frac{E_\omega}{|\vec{p}_\omega|} \Gamma_{abs}
\end{equation}

\noindent 
and the corresponding survival probability is given by
\begin{equation}
\label{EF}
 \exp\left[ 
\int_0^{\infty} dl 
(-1) \frac{E_\omega}{|\vec{p}_\omega|} 
\Gamma_{abs}\Big(\rho(\vec{r}\, ')\Big)\right],
\end{equation}

\noindent where
 $\vec{r}\,'=\vec{r}+l\frac{\vec p_\omega}{|\vec p_\omega|}$
 with  $\vec{r}$ being the $\omega$ production point inside the
nucleus. 


Then the total photonuclear 
cross section $A(\gamma,\omega)X$ is given 
by Eq.~(\ref{TCSnuclear}) where now 
the kernel $\Omega_{\omega}$ is replaced by the eikonal factor of
Eq.~(\ref{EF}).
Since, the integration over $\widetilde{m}_{\omega}^2$ in
Eq.~(\ref{TCSnuclear}) should not change
the normalization of the total cross section,
one can use in Eq.~(\ref{TCSnuclear}) 
the spectral function of the free $\omega$,
$S_{\omega} \to S^{free}_{\omega}$. 

\begin{figure*}[t]
\begin{center}
\includegraphics[clip=true,width=0.49\columnwidth,angle=0.]
{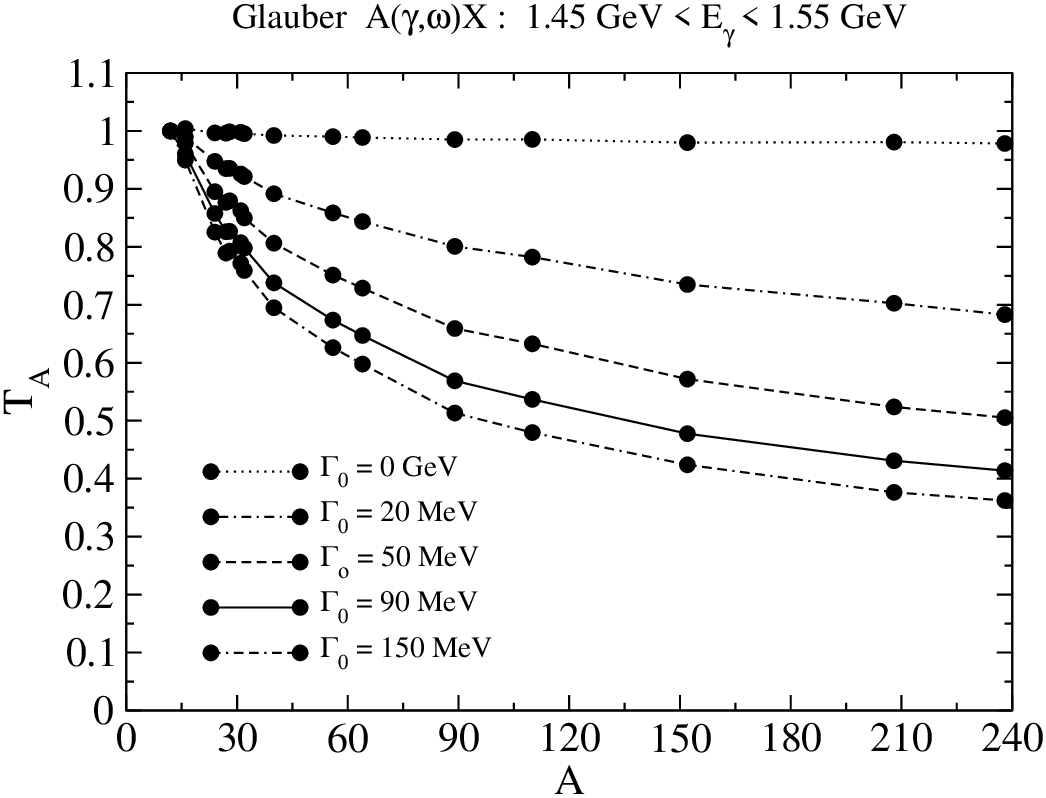}
\includegraphics[clip=true,width=0.49\columnwidth,angle=0.]
{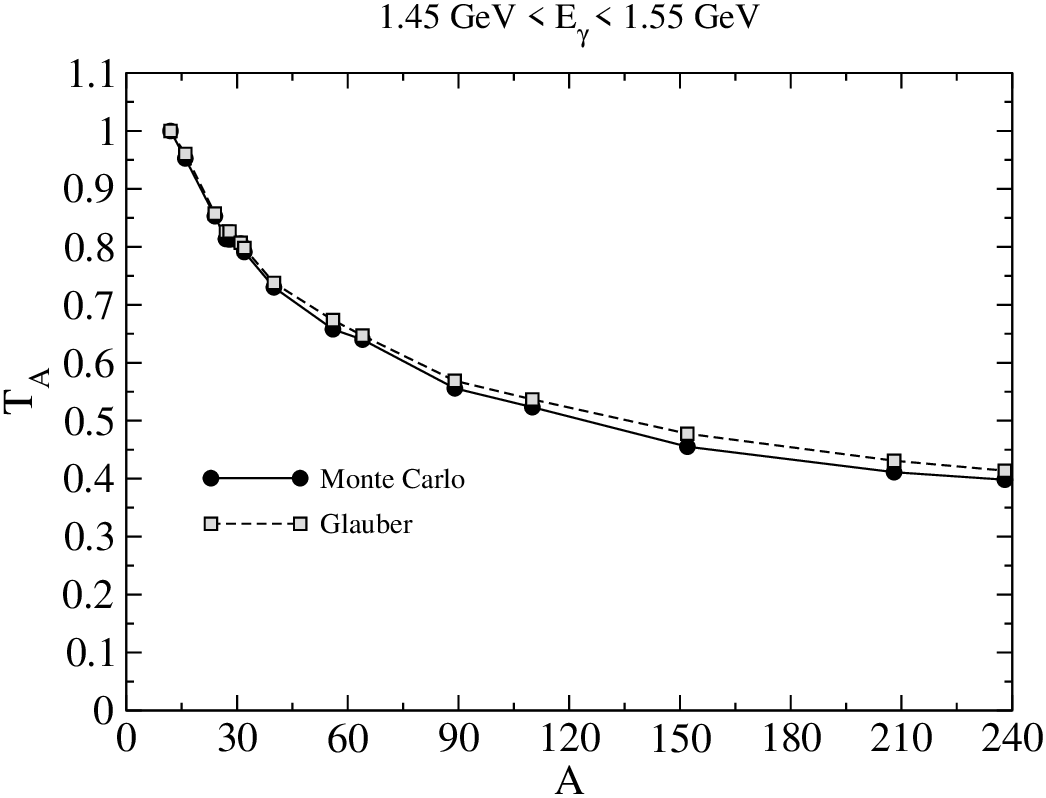}
\caption{\label{FigTrEk}  \footnotesize
Left panel: The $A$-dependence
of the nuclear transparency ratio $T_A$ when
using the eikonal approximation (see the text).
The notations for the curves are the same as in Fig.~\ref{FigTrMC}.
Right panel: Comparison of the $A$-dependence
of the nuclear transparency ratio calculated 
using the Monte Carlo simulation method (solid curve) and the eikonal
approximation (dashed curve). The inelastic width of the $\omega$
of $\Gamma_0 = 90$~MeV at $\rho_0$ has been assumed in both calculations.
}
\end{center}
\end{figure*}

The results of the eikonal approximation
for the $A$-dependence
of the nuclear transparency ratio $T_A$ are shown in Fig.~\ref{FigTrEk}
(left panel).
The kinematic constraints and notations for the curves are the same as in
Fig.~\ref{FigTrMC}. In Fig.~\ref{FigTrEk} (right panel)
we compare the $A$-dependence
of the nuclear transparency ratio calculated
using the Monte Carlo simulation method (solid curve) and the eikonal
approximation (dashed curve).  
In both cases we used
the collisional width (absorptive part) of 90~MeV at 
normal nuclear matter densities $\rho_0$ which gives a fair description
of the preliminary data of the CB/TAPS
collaboration~\cite{Kotulla:2006wz}. 
As one can see,  the two curves are very close to each other suggesting
a remarkable accuracy of the eikonal approximation. See also related
discussions in Ref.~\cite{muhlich2}.

There are essential differences in the two approaches. In the eikonal 
approximation the $\omega$ proceeds always in the forward direction after
any collision following the straight trajectories in their way out 
of the nucleus. Furthermore, in the eikonal 
approximation the $\omega$-mesons  keep always their original energy, which is
not the case in the MC simulation, where because of the quasielastic steps 
the energy depends on the scattering
angle. Also FSI of the pions coming from the
interior of the nucleus is not accounted for in the eikonal formula.
In spite of that, the results of both methods are rather
similar. Although the eikonal  method is very accurate 
for the total cross sections 
at rather big values of the absorption parameter $\Gamma_0$, 
it cannot be used for the detailed studies of the differential spectra
where the acceptance conditions relevant for the actual experiment
must be taken into account like the MC simulation method does. 

Finally, using the results of both methods and taking into account
the preliminary results of CBELSA/TAPS experiment~\cite{Kotulla:2006wz} 
we get an estimate 
for the $\omega$ width
\begin{equation}
\Gamma_{abs} \simeq 90 \times \frac{\rho(r)}{\rho_0}~\mbox{MeV}.
\end{equation}
This estimate must be understood as an average over the 
$\omega$ three momentum. 
By this we conclude that
the measurements of $A$ dependence of the nuclear transparency ratio
provide very important information on
the absorptive part of the $\omega$-meson  width 
in the nuclear medium.

\begin{figure}[t]
\begin{center}
\includegraphics[clip=true,width=0.5\columnwidth,angle=0.]
{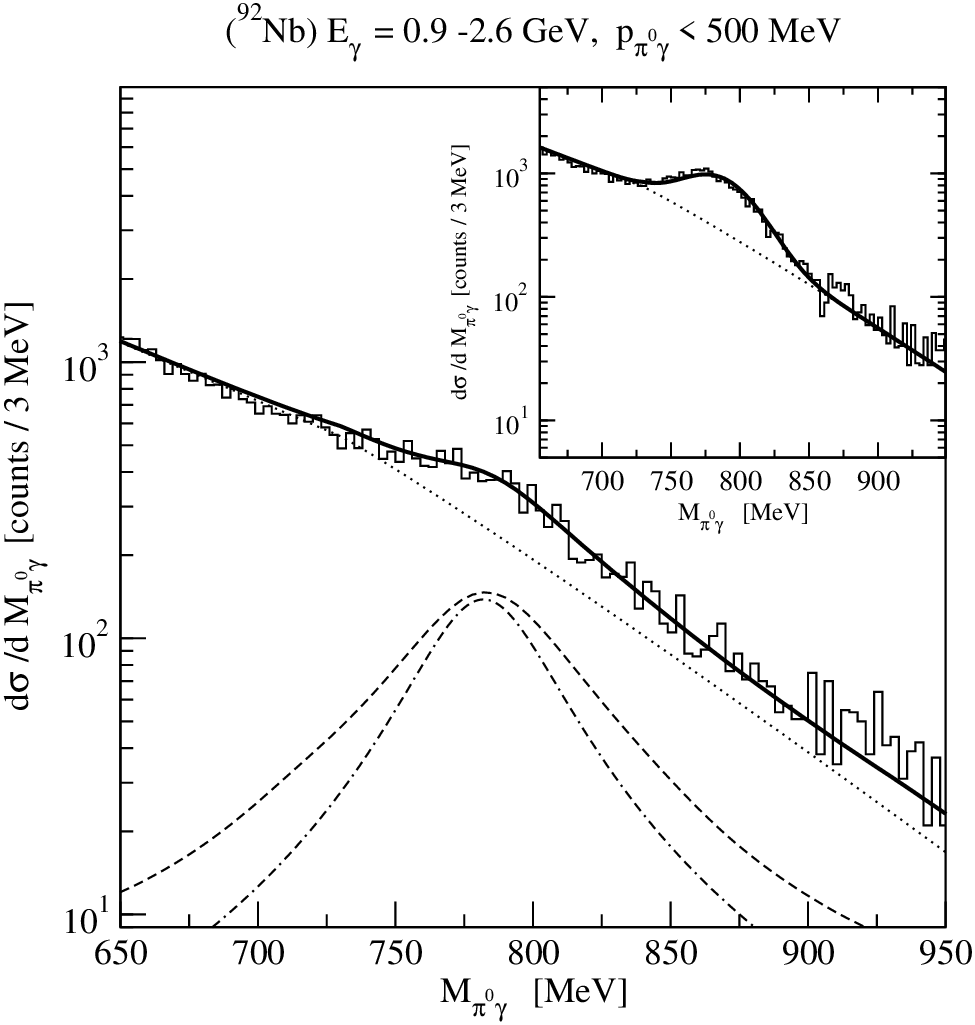}
\caption{\label{PRL}  \footnotesize
Invariant mass spectra reconstructed from the $\pi^0\gamma$
events in the $(\gamma,\pi^0\gamma)$ reaction from Nb target. 
The experimental data are from Ref.~\cite{trnka}. 
Dotted curve is an uncorrelated $\pi^0\gamma$
background (see the text). 
The dashed and dash-dash-dotted curves correspond to the
$\omega \to \pi^0\gamma$ events with and without the kinematic cut
$|\vec{p}_{\pi^0\gamma}| < 500$ MeV,
respectively. The normalization without cut is arbitrary.
The solid line corresponds to the sum of the
background and the dashed line.
Inset: The $\pi^0\gamma$ invariant mass spectra in the elementary
$p(\gamma,\pi^0\gamma)p$ reaction. 
Same background (dotted curve) as for the Nb target has been used.
The solid line is the sum of the
background and $\omega \to \pi^0\gamma$ events.}
\end{center}
\end{figure}


\begin{figure}[t]
\begin{center}
\includegraphics[clip=true,width=0.5\columnwidth,angle=0.]
{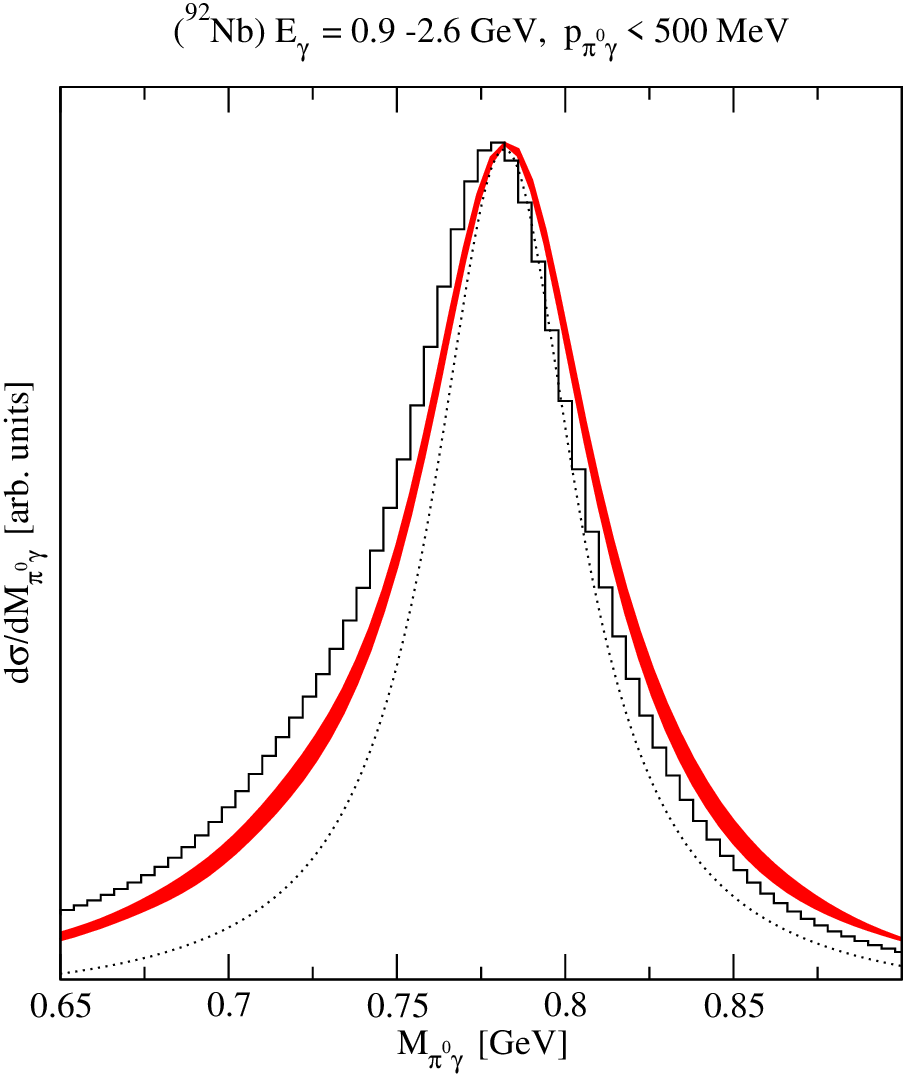}
\caption{\label{PRL2} \footnotesize
Invariant mass spectra reconstructed from $\pi^0\gamma$ events.
The kinematic constraints are the same as in Fig.~\ref{PRL}.
The band corresponds to the changes of the in--medium $\omega$-mass
according to $m_{\omega}\pm 40 (\rho/\rho_0) $~MeV. The serrated line 
corresponds
to the scaling of the $\omega$ mass $m_{\omega}(1- 0.16 \rho/\rho_0)$.
The dotted curve is the $\omega\to \pi^0\gamma$ signal without
applying the kinematic cuts. See text for further explanations.
}
\end{center}
\end{figure}

\section{In-medium $\omega$-meson mass and CBELSA/TAPS experiment}

As we have already seen in the previous sections 
the $\omega$-mesons which have an increased decay probability 
inside the nucleus may carry experimentally observable
information concerning their in-medium properties.
Furthermore, the $\omega$-mesons decaying inside nuclei can be used for  
studying their in-medium properties. 
Success in finding experimental information on these properties suggests
to take a certain kinematic condition where the
decay length $L_{\omega} = |\vec{p}_{\omega}|/m_{\omega} \Gamma_{\omega}$
of the $\omega$ moving with the three momentum $|\vec{p}_{\omega}|$
should be less than the nuclear radius.
Therefore, in the actual experiment
it is preferred that the kinetic energy of the $\omega$ meson 
is small, since in this case,
the fraction of $\omega$ mesons decaying inside the nucleus can be
increased further merely by minimizing the $\omega$ decay length.
This can be achieved with an incident photon energy close to
the $\omega$-production threshold  or using
the kinematical cuts on the $\omega$-meson three momentum,
as have been demonstrated 
in Refs.~\cite{Messchendorp:2001pa,muhlich}.
The later idea to gate the $\omega$ momenta using the higher momentum cuts
has been used in a recent CBELSA/TAPS experiment where 
a significant modification of the $\omega$-meson line shape 
when these mesons are produced in a dense medium was reported~\cite{trnka}. 
This change of the $\pi^0\gamma$ invariant mass spectra and 
an accumulation of additional
strength at lower invariant masses was interpreted as
an evidence for the lowering of the $\omega$ mass in nuclei. 

At the same time one should note, that the $\omega$ line shape reconstructed
from $\pi^0\gamma$ events strongly depends
on the background shape subtracted from the bare $\pi^0\gamma$ signal.
In Ref.~\cite{trnka} the shape of the background was chosen such that it
accounted for all the experimental strength at large invariant masses.
This choice was done both for the elementary $\gamma p \to  \pi^0 \gamma p$
reaction as well as for nuclei. As we shall show, this choice of background in
nuclei implies a change of the shape from the elementary reaction to that in
the nucleus for which no justification was given. We shall also show
that when the
same shape for the background as for the elementary reaction is chosen, 
the experiment in nuclei shows strength at invariant masses higher than 
$m_{\omega}$ where the choice of~\cite{trnka} necessarily produced no
strength. We will also see that the experimental data can be naturally
interpreted in terms of the large in-medium $\omega$ width discussed above
without
the need to invoke a shift in the $\omega$ mass in the medium.


In Fig.~\ref{PRL}  
we show the experimental data (solid histogram) for the 
$\pi^0\gamma$ invariant mass spectra in the reaction
$(\gamma,\pi^0\gamma)$~\cite{trnka} from $^{92}_{41}$Nb target. 
The insert corresponds to the $\pi^0\gamma$ spectra from the
hydrogen target. 
In our MC calculations the incident photon beam has been constrained
in the range $0.9$~GeV $< E_{\gamma}^{in} <$ $2.6$~GeV. 
The higher momentum cut 
$|\vec{p}_{\pi^0\gamma}| = |\vec{p}_{\pi^0}+\vec{p}_{\gamma}|< 500$~MeV
on a three momentum of the $\pi^0 \gamma$ pair
was imposed as in the actual experiment. First, we use the hydrogen target,
see insert in Fig.~\ref{PRL}, 
to fix the contribution of the uncorrelated $\pi^0\gamma$
background  (dotted curve) which together with
the $\pi^0\gamma$ signal from $\omega \to \pi^0\gamma$ decay,
folded with the Gaussian experimental resolution of 55 MeV as 
in Ref.~\cite{trnka}, gives a fair
reproduction of the experimental spectra. Then we assume the same 
shape of the $\pi^0\gamma$ 
background in the photonuclear reaction. 
In the following we use the $\omega$ inelastic 
width of $\Gamma_0 = 90$~MeV
at $\rho_0$.
The exclusive
$\omega \to \pi^0\gamma$ MC spectra is shown by the dashed curve. 
The solid curve is the reconstructed
$\pi^0\gamma$ signal after applying the cut on $\pi^0\gamma$ momenta and
adding
the background fixed when using the hydrogen target (dotted curve). 
Note that the shape of the exclusive $\pi^0 \gamma$ signal without applying 
a cut on $\pi^0\gamma$ momenta
(dash-dotted curve) is dominated by the experimental resolution 
and no broadening of the $\omega$ is observed. This is in agreement
with data of Ref.~\cite{trnka}.
But applying the cut one increases
the fraction of in-medium decays coming from the interior of the nucleus 
where the spectral function is rather broad
and as a result the broadening of the $\pi^0\gamma$ signal with respect
to the signal (without cut) can be well seen.
The resulting MC spectra (solid curve) 
shows the accumulation of the $\pi^0\gamma$ events from the left and right 
sides
of the mass spectra, and it is consistent both with 
our choice of the uncorrelated $\pi^0\gamma$
background and experimental data.
 
It is interesting to stress here that the choice of the background done 
in~\cite{trnka} significantly changes it from the proton target to 
the nucleus.
However, inspection of Fig.~\ref{PRL} (a) of Ref.~\cite{trnka} clearly shows
that
while the background on the proton has a kind of convex parabolic form
(see this also in Fig.~\ref{PRL} here, insert), the background
chosen for Nb in~\cite{trnka} is a straight line in the logarithmic plot.
This increases the assumed background with respect to that induced from the
proton experiment at high masses which is difficult to justify. 
Indeed, even if the  distortion on the pions from FSI is not large, its effect 
should go into degrading the pion energy and consequently moving events to
lower $\pi^0\gamma$ invariant masses, hence reducing relatively the background
at higher $\pi^0\gamma$ masses, not increasing it.

We have also done the exercise of seeing the sensitivity of the results to 
changes in the mass. In the dark band of Fig.~\ref{PRL2} we show the results of
having the $\omega$ mass in between $m_{\omega}\pm 40 \rho/\rho_0$~MeV.
The narrowness of the band indicates that the experimental data could not be 
precise enough to distinguish between these cases. In other words, this 
experiment is too insensitive to changes in the mass to be used for a precise
determination of the shift of the $\omega$-mass in the nuclear medium.
On the other hand, we have shown that the results are more sensitive to 
changes in the width of the $\omega$ in the medium. This is due to the fact
that an increased width produces more decays of the $\omega$ in the medium
which allows one to see changes in the $\pi^0\gamma$ spectra. This larger 
sensitivity of the results to the width than to mass change was already
observed in studies of the $\rho$ production in  nuclei~\cite{Oset:2001iy}.
In any case, we have checked what would be the results should we 
assume $m_{\omega}(1-0.16 \rho/\rho_0)$, i.e., a shift of 125~MeV at 
$\rho = \rho_0$
as suggested in Ref.~\cite{hatsuda}. We can see the results in the  
serrated line in Fig.~\ref{PRL2} which shows a visible asymmetry with respect
to the thick solid line, the one we have shown to be compatible with 
experiment. We should note that the peak position does not move since it is
dominated
by the decay of the $\omega$ outside the nucleus.
 
\section{\label{Summary} Conclusions}

We have performed calculations of the $(\gamma ~ A \to \pi^0 \gamma ~X)$
 reaction for $\pi^0 \gamma$ invariant masses around the $\omega$ mass. A clear
 signal is seen in the experiment for $\omega$ production with subsequent 
$\pi^0 \gamma$ decay, both for the elementary reaction on a proton target as
well as in the nuclear targets.  In addition there is a sizable background that
has to be subtracted in order to identify the $\omega$ signal in the 
proton and nuclear targets.  We have performed a Monte Carlo simulation of the
$\omega$ production followed by its decay into $\pi^0 \gamma$, with a detailed
study of the final state interaction of the particles involved in the process.
The invariant mass distribution of the $\pi^0 \gamma$ pairs is evaluated when
the particles have left the nucleus.  We compare our theoretical results with
the experimental data and induce from there that the data are compatible 
with an $\omega$ in the medium width of around 90 MeV at normal nuclear matter
and no shift in the mass.  This large width in the medium is compatible with
preliminary results obtained from the transparency ratio in $\omega$ 
production in
nuclei for which we also present theoretical results here. 

   The results obtained here for the mass shift disagree with those formerly
claimed in \cite{trnka} and we show that the reason
for this discrepancy  is due to a different choice of background. In 
\cite{trnka} the background for nuclear targets did not scale with respect to
the one on proton targets and the shape assumed for both targets was 
manifestly different.  The choice of background in nuclei was done such 
as to cut the
contribution of high $\omega$ invariant masses. By means of this choice, the
shape of the $\omega$ mass distribution in nuclei was asymmetric, showing
additional strength only at masses smaller than the $\omega$ mass which induced
the authors to claim that there was a shift of mass to lower invariant masses.
  We have done a different choice of background, more suited for the studied
reaction,
which is to take the same shape for the background in nuclei as for the proton
target.  
 With this choice of background, the $\omega$ invariant mass distribution is
 symmetric and  explained  in terms of the enlarged $\omega$ width in the
 medium with no need to invoke a shift in the mass. We also show that the 
reaction is not well suited to make precise determinations of the mass, and
see that the data could not distinguish between masses in a range
$m_{\omega}\pm 40 \rho/\rho_0$~MeV. 
 
\section*{Acknowledgments}  
We would like to acknowledge useful discussions with V. Metag, 
M. Kotulla and D. Trnka as well as their help in providing their data.
This work is partly supported by DGICYT contract
number BFM2003-00856, the Generalitat Valenciana, the projects FPA2004-05616
(DGICYT) and SA104/04 (Junta de Castilla y Leon) 
and the E.U. EURIDICE network
contract no.  HPRN-CT-2002-00311. This research is  part of the EU Integrated
Infrastructure Initiative  Hadron Physics Project under  contract number
RII3-CT-2004-506078.

\end{document}